\documentclass{eptcs}

\usepackage{iftex}

\ifpdf
  \usepackage{underscore}         
  \usepackage[T1]{fontenc}        
\else
  \usepackage{breakurl}           
\fi

\usepackage{url}
\usepackage{booktabs}
\usepackage{graphicx}
\usepackage{stmaryrd}
\usepackage{xcolor}
\usepackage{xargs}
\usepackage{textcomp}
\usepackage{amsmath}
\usepackage{amssymb}
\usepackage{relsize}
\usepackage{xspace}
\usepackage{amsfonts}
\usepackage{mathtools}
\usepackage{framed}
\usepackage{subfig}
\usepackage{algorithmic}

\newcommand{\figpath}{.}

\newtheorem{example}{Example} 

\DeclareMathOperator*{\argmin}{arg\,min}

\newcommand{\lbr}{\mbox{$\lbrack\hspace{-0.3ex}\lbrack$}}
\newcommand{\rbr}{\mbox{$\rbrack\hspace{-0.3ex}\rbrack$}}
\newcommand{\eval}[2]{\lbr #1 \rbr_{#2}}

\usepackage{tikz}
\usetikzlibrary{shapes,arrows,chains}
\usepackage{verbatim}
\usepackage[titlenumbered,ruled,linesnumbered]{algorithm2e}
\usepackage{listings}
\newcommand{\prettylstciao}[0]{
\lstset{language=Prolog,
  frameround=tttt,
  xleftmargin=0.2cm,
  rulecolor=\color{blue},
  numbers=left,numberstyle=\tiny,stepnumber=1,numbersep=8pt,
  tabsize=4,
  showstringspaces=false,
  breaklines=true,breakatwhitespace=true,
  showlines=true,
  showspaces=false,showtabs=false,
  upquote=true,
  commentstyle=\color{gray},
  keywordstyle=\color{eminence},
  basicstyle=\small\ttfamily,
  keywordstyle=\color{weborange},
  emphstyle={\color{blue}},
  emph={pred,prop,trust,check,checked,true,rsize,cardinality,
  not_fails,module,exp,cost,steps_ub,steps_lb,size_ub,size_lb,
  covered,mut_exclusive,cost,use_module,int,calls,success,cost_center,
  is_det,num,nat,var,list,ground,length,terminates,term,steps_o,resource,
  entry,impl_defined,regtype},
  otherkeywords={>,<,>=,=<,.,;,-,!,=,~,*,\&,+,:-,[,],|,->,:,:=},
  morekeywords= {>,<,>=,=<,.,;,-,!,=,~,*,\&,+,:-,[,],|,->,:,:=},
  escapeinside=\`\`,
}}

\newcommand{\ciaopp}{CiaoPP\xspace}

\definecolor{eminence}{RGB}{108,48,130}
\definecolor{weborange}{RGB}{200,130,0}

\newcommand\stdub[2]{{\tt C}_{#1}(#2)}
\newcommand\sizeub[2]{{\tt S}_{#1}(#2)}

\newcommand\costrelations{cost relations}

\pgfdeclarelayer{marx}
\pgfsetlayers{main,marx}
\providecommand{\cmark}[2][]{%
  \begin{pgfonlayer}{marx}
    \node [nmark] at (c#2#1) {#2};
  \end{pgfonlayer}{marx}
  } 
\providecommand{\cmark}[2][]{\relax} 

\title{Solving Recurrence Relations using Machine Learning, with
  Application to Cost Analysis}
  
\def\titlerunning{Solving Recurrence Relations using Machine Learning}
\def\authorrunning{M. Klemen, M.\'{A}. Carreira-Perpi\~{n}\'{a}n \& P. Lopez-Garcia}

\author{$^{1,2}$Maximiliano Klemen,
        $^{3}$Miguel \'{A}. Carreira-Perpi\~{n}\'{a}n, and
        $^{4,2}$Pedro Lopez-Garcia
  \vspace{1mm}
  \institute{$^{1}$Universidad Polit\'ecnica de Madrid (UPM) Madrid, Spain}
  \institute{$^{2}$IMDEA Software Institute, Madrid, Spain}
  \institute{$^{3}$University of California, Merced, USA}
  \institute{$^{4}$Spanish Council for Scientific Research, Madrid, Spain}
  \email{\{maximiliano.klemen,pedro.lopez\}@imdea.org, mcarreira-perpinan@ucmerced.edu}
}

\hypersetup{
  bookmarksnumbered,
  pdftitle    = {\titlerunning},
  pdfauthor   = {\authorrunning},
  pdfsubject  = {EPTCS},   
  pdfkeywords = {Cost Analysis, Recurrence Relations, Static Analysis, Machine Learning,Sparse Linear Regression,
                 Resource Usage Analysis} 
}

\begin{document}
\maketitle   

\begin{abstract}
Automatic static cost analysis infers information about the resources
used by programs without actually running them with concrete data, and
presents such information as functions of input data sizes.  Most of
the analysis tools for logic programs (and other languages) are based
on setting up recurrence relations representing (bounds on) the
computational cost of predicates, and solving them to find closed-form
functions that are equivalent to (or a bound on) them. Such recurrence
solving is a bottleneck in current tools: many of the recurrences that
arise during the analysis cannot be solved with current solvers, such
as Computer Algebra Systems (CASs), so that specific methods for
different classes of recurrences need to be developed.  We address
such a challenge by developing a novel, general approach for solving
arbitrary, constrained recurrence relations, that uses
machine-learning sparse regression techniques to \emph{guess} a
candidate closed-form function, and a combination of an SMT-solver and
a CAS to \emph{check} whether such function is actually a solution of
the recurrence. We have implemented a prototype and evaluated it with
recurrences generated by a cost analysis system (the one in
\ciaopp). The experimental results are quite promising, showing that
our approach can find closed-form solutions, in a reasonable time, for
classes of recurrences that cannot be solved by such a system, nor by
current CASs.
\end{abstract}


\section{Introduction and Motivation}


The motivation of the work presented in this paper stems from
automatic static cost analysis and verification of logic
programs~\cite{granularity-short,caslog-short,low-bounds-ilps97-short,resource-iclp07-short,plai-resources-iclp14-short,gen-staticprofiling-iclp16-short,resource-verification-tplp18-shortest}.
The goal of
such analysis is to infer information about the resources used by
programs without actually running them with concrete data, and present
such information as functions of input data sizes and possibly other
(environmental) parameters.
%
We assume a broad concept of resource as a numerical property of the
execution of a program, such as number of \emph{resolution steps},
\emph{execution time}, \emph{energy consumption}, \emph{memory},
number of \emph{calls} to a predicate,
number of \emph{transactions} in a database, etc.
%
%
Estimating in advance the resource usage of computations is useful for
a number of applications, such as
automatic program optimization,
verification of resource-related specifications, detection of
performance bugs, helping developers make resource-related design
decisions,
security applications (e.g., detection of side
channels attacks),
or blockchain platforms (e.g., smart-contract gas analysis and
verification).


The challenge we address originates from the established approach of
setting up recurrence relations representing the cost of predicates,
parameterized by input data
sizes~\cite{Wegbreit75,Rosendahl89,granularity-short,caslog-short,low-bounds-ilps97-short,resource-iclp07-short,AlbertAGP11a-short,plai-resources-iclp14-short,gen-staticprofiling-iclp16-short},
which are then solved to obtain \emph{closed forms} of such
recurrences (i.e., functions that provide either exact, or upper/lower
bounds on resource usage in general).
Such approach can infer different classes of functions (e.g.,
polynomial, factorial, exponential, summation, or logarithmic).


The applicability of these resource analysis techniques strongly
depends on the capabilities of the component in charge of solving
(or safely approximating)
the recurrence relations generated during the
analysis, which has
become a bottleneck in some systems.


A common approach to automatically solving such
recurrence relations
consists of
using a Computer Algebra System (CAS) or a specialized solver to find
a closed form. However, this approach poses several difficulties and
limitations.
For example,
some recurrence relations contain complex expressions or recursive
structures that most of the well-known CASs cannot solve, making
it necessary to develop ad-hoc techniques to handle such cases.
Moreover, some recurrences may not have the form required by such
systems because an input data size variable does not decrease, but
increases instead. Note that a decreasing-size variable could be
implicit in the program, i.e., it could
be a function of a subset input data sizes (a ranking function), which
could be inferred by applying established techniques
used in termination analysis~\cite{PodelskiR04b-short}. However, such
techniques are usually restricted to linear arithmetic.

\begin{figure}[tb!]
\centering
\includegraphics[scale=0.43]{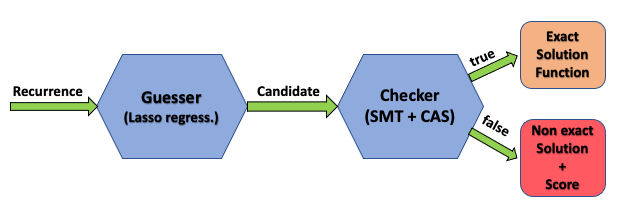}
\caption{Control flow diagram of our novel solver based on machine learning.} \label{fig1}
\end{figure}

In order to address this challenge
we have developed a novel, general method for solving arbitrary,
constrained recurrence relations.
It is a \emph{guess and check} approach that uses
machine learning techniques for the \emph{guess} stage, and a
combination of an SMT-solver and a CAS for the
\emph{check} stage (see Figure~\ref{fig1}).
To the best of our knowledge, there is no
other approach that does this.
The resulting closed-form function solutions can be of different
kinds, such as polynomial, factorial, exponential, summation, or
logarithmic.

The rest of this paper is organized as follows.
Section~\ref{sec:overview} gives and overview of our novel \emph{guess
and check} approach. Then Section~\ref{sec:prelim} provides some
background information and preliminary notation.
Section~\ref{sec:description-approach} presents a more detailed,
formal and algorithmic description of our approach.
Section~\ref{sec:application-cost-analysis} describes the use of our
approach in the context of
static cost analysis. Section~\ref{sec:implem-and-eval} comments on our
prototype implementation and its experimental evaluation.  Finally,
Section~\ref{sec:conclusions} summarizes some conclusions and lines
for future work.

\section{Overview of our Approach}
\label{sec:overview} 

We now give an overview of the two stages of our approach already
mentioned: \emph{guess} a candidate closed-form function, and
\emph{check} whether such function is actually a solution of the
recurrence relation.

Given a recurrence relation for a function $f(\vec{x})$,
solving it means to find a closed-form function $\hat{f}(\vec{x})$
that has the same domain as $f(\vec{x})$, and for all $\vec{x}$ in such domain,
$\hat{f}(\vec{x}) = f(\vec{x})$.
By a closed-form function $\hat{f}$ we mean an expression that is built
by using only elementary arithmetic functions, e.g., constants,
addition, subtraction, multiplication, division, exponential, or even
factorial functions. In particular, this means that $\hat{f}$ does not
contain any subexpressions built by using the same function $\hat{f}$
(i.e., $\hat{f}$ is not recursively defined).
We will use the following recurrence as an example to illustrate our approach:
\begin{equation}
\label{eq:recurrence}
  \begin{array}{rll}
    f(x) = & 0 &  \text{ if } x = 0 \\ [-1mm]
    f(x) = & f(f(x-1)) + 1 &  \text{ if } x > 0 \\ [-1mm]
  \end{array}
\end{equation}
\noindent 

\subsection{The ``guess'' stage (sparse linear regression via Lasso)}

We use a sparse linear regression mechanism (see Section~\ref{sec:prelim}
for more details), so that any possible model we can obtain (which
constitutes a candidate solution) must be a linear combination of a
predefined set of terms, but using a usually small subset of terms.
That is, a
function $\hat{f}(\vec{x})$ of the form:
\begin{equation*}
 \hat{f}(\vec{x}) = \beta_0 + \beta_1 \ t_1(\vec{x}) + \beta_2 \ t_2(\vec{x}) + \cdots + \beta_n \  t_n(\vec{x})  
\end{equation*}
\noindent
where the $t_i$'s are arbitrary functions on $\vec{x}$ from a set $T$ of
candidate terms that we call \emph{base functions}, and the $\beta_i$'s
are the coefficients (real numbers) that are estimated by regression, but
so that only a few coefficients are nonzero.
Currently, the set $T$ is fixed, and contains the base functions that
are representative of the
common complexity orders (in Section~\ref{sec:conclusions} we comment on
future plans to obtain it).  For illustration purposes, assume that we
use the following set $T$ of
base functions:
\begin{equation*}
    \begin{array}{rll} T = \{ \lambda x.x, \lambda x.x^2,
      \lambda x.x^3, \lambda x.\lceil \log_2(x) \rceil, \lambda x. 2^x, 
\lambda x.x \cdot \lceil \log_2(x) \rceil \}
    \end{array} 
\end{equation*}
\noindent
where each base function is represented as a lambda expression.  Then,
the sparse linear regression is performed as
follows:
\begin{enumerate}
\item Generate a training set $S$.
First, a set $X_{\text{train}} = \{ \vec{x}_1, \ldots,\vec{x}_k \}$ of input
values to the recurrence function is randomly generated.  Then,
starting with an initial $S = \emptyset$, for each input value
$\vec{x}_i \in X_{\text{train}}$, a training case $s_i$ is generated and
added to $S$. 
For any input value $\vec{x} \in X_{\text{train}}$ the corresponding training
case $s$ is a tuple of the form:
\begin{equation*}
s = \langle b, c_1, \ldots, c_n \rangle
\end{equation*}
\noindent
where $c_i = \eval{t_i}{\vec{x}}$ for $1 \leq i \leq n$, and
$\eval{t_i}{\vec{x}}$ represents the result (a scalar) of evaluating
the base function $t_i \in T$ for input value $\vec{x}$, where $T$ is
a set of $n$ base functions, as already explained. The (dependent)
value $b$ (also a constant number) is the result of evaluating the
recurrence $f(\vec{x})$ that we want to solve or approximate, in our
example, the one defined in Equation~\ref{eq:recurrence}.
Assuming that there is an $\vec{x} \in X_{\text{train}}$ such that $\vec{x} =
\langle 5 \rangle$, its corresponding training case $s$ in our example
will be:
\begin{equation*}
   \begin{array}{rll}
     s &=& \langle
     \mathbf{f(5)},
     \eval{x}{5}, \eval{x^2}{5}, \eval{x^3}{5}, \eval{\lceil \log_2(x) \rceil}{5},
     \ldots \rangle \\ &=& \langle \mathbf{5}, 5, 25, 125, 3, \ldots
     \rangle \\
   \end{array} 
\end{equation*}
\item Perform the sparse regression in two steps using the training set $S$
  created above.  In the first step, we use linear regression with
  Lasso ($\ell_1$) regularization~\cite{Hastie-15a} on the coefficients.
  This is a penalty term that encourages coefficients whose associated base
  functions have a small correlation with the
  dependent value to be exactly zero. This way, typically most
  of the
  base functions
  in $T$ will be discarded, and only those that
  are really needed to approximate our target function will be kept. The level of
  penalization is controlled by a hyperparameter
  $\lambda \ge 0$. As commonly done in machine learning~\cite{Hastie-15a},
  the value of $\lambda$ that generalizes optimally on unseen (test) inputs is
  found via cross-validation on a separate validation set (generated randomly
  in the same way as the training set).
The result of this step is a (column) vector $\vec{\beta}$ of
coefficients,
and an independent coefficient $\beta_0$. Finally, we generate a test set
$X_{\text{test}}$ (again, randomly in the same way as the training set) of
input values to the recurrence function
to obtain a measure $R^2$ of the accuracy of
the estimation. Additionally, we discard those terms whose
corresponding coefficient is less than a given threshold $\epsilon$.
The resulting closed-form expression that estimates the target
function is
\begin{equation*}
  \begin{array}{rll}
   \hat{f}(\vec{x}) = \mathsf{rm}_{\epsilon}(\vec{\beta}^T) \cdot E(T,\vec{x}) + \beta_0
  \end{array} 
\end{equation*}
where $E(T,\vec{x})$ is a vector of the terms in $T$ with the
arguments bound to $\vec{x}$, and $\mathsf{rm}_{\epsilon}$ takes a vector of
coefficients and returns another vector where the coefficients
less than $\epsilon$ are rounded to zero.  
Both the Lasso regularization and the pruning function
discard
many terms from $T$ in the final
function.

\item Finally, our method
performs again a standard linear regression (without Lasso
regularization) on the training set $S$, but without using those base functions
corresponding to the terms discarded previously by Lasso and the
$\epsilon$-pruning. In our example, with $\epsilon = 0.05$,
we obtain:
\begin{equation*}
  \begin{array}{rl}
    \hat{f}(x) = 1.0 \ x & 
  \end{array} 
\end{equation*}
with a value $R^2 = 1$, which means that the estimation obtained
predicts exactly the values for the test set, and thus, it is a
candidate solution for the recurrence in Equation~\ref{eq:recurrence}.
If $R^2$ were less than $1$, it would mean that
the function obtained is not a candidate (exact) solution, but a
(possibly unsafe) approximation, as there are values in the test set that cannot be
exactly predicted.

\end{enumerate}

\subsection{The ``check'' stage}

Once a function that is a candidate solution for the recurrence has
been guessed, the second step of our method
tries to verify whether such a candidate is actually a solution. To do
so, 
the recurrence
is encoded as a first order
logic formula
where the references to the target function are replaced by the
candidate solution whenever possible. Afterwards, we use an
SMT-solver to check whether
the negation of such formula is satisfiable, in which case we can
conclude that the candidate is not a solution for the
recurrence. Otherwise, if such formula is unsatisfiable, then the
candidate function is
an exact solution. Sometimes, it is necessary
to consider a precondition for the domain of the recurrence, which is
also included in the encoding.

To illustrate this process,
Expression~(\ref{eq:verexample1-rec}) below shows the recurrence
relation we target to solve, followed by the candidate solution
obtained previously using linear regression:
\begin{equation}
  \label{eq:verexample1-rec}
  \begin{array}{rll}
    f(x) = & 0 &  \text{ if } x = 0 \\ [-1mm]
    f(x) = & f(f(x-1))  + 1 &  \text{ if } x > 0 \\ [-0mm]
    \hat{f}(x) = & x &  \text{ if } x \geq 0 \\ [-1mm]
  \end{array}
\end{equation}
\noindent
Now, Expression~(\ref{eq:verexample1-form}) below shows the encoding
of the recurrence as a first order logic formula.
\begin{equation}
  \label{eq:verexample1-form}
  \forall x \: \left( (x = 0 \implies \underline{f(x)} = 0)  \wedge (x>0 \implies \underline{f(x)} =
  \underline{f(\underline{f(x-1)})} + 1) \right)
\end{equation}
\noindent
Finally, Expression~(\ref{eq:verexample1-negform}) below shows the negation
of such formula, as well as the references to the function name
substituted by the definition of the candidate solution. We underline
both the subexpressions to be replaced, and the subexpressions
resulting from the substitutions.
\begin{equation}
  \label{eq:verexample1-negform}
  \exists x \: \neg (\left( (x = 0 \implies \underline{x} = 0)  \wedge (x>0 \implies \underline{x} = \underline{x-1} + 1) \right))
\end{equation}
\noindent
It is easy to see that Formula~(\ref{eq:verexample1-negform})
is unsatisfiable. Therefore, $\hat{f}(x) = x$ is an 
exact solution for $f(x)$ in the recurrence defined by
Equation~\ref{eq:recurrence}.

For some cases where the candidate solution contains transcendental
functions, our
implementation of the method uses
a CAS to perform simplifications and
transformations, in order to obtain a formula supported by
the SMT-solver. We find this combination of CAS and SMT-solver
particularly useful, since it allows solving more problems than only
using one of these systems in isolation.

\section{Preliminaries}
\label{sec:prelim}

\paragraph{\textbf{Recurrence relations}.}
A recurrence relation of order
$k$, $k > 0$, for a function $f$, is a set of equations that give $k$
initial values for $f$, and an equation that recursively defines any
other value of $f$ as a function $g$ that takes $k$ previous values of
$f$ as parameters. For example, the following recurrence relation of
second order ($k=2$), with $g$ being the arithmetic addition $+$,
defines the Fibonacci function:
\begin{equation}
\label{eq:fibo-recurrence}
f(n) = 
  \begin{cases*}  
    1 &  \text{ if } n = 0 \text{ or } n = 1 \\ 
    f(n-1) + f(n-2) &  \text{ if } n $\geq$ 2 \\ 
  \end{cases*}
\end{equation}
A challenging class of recurrences 
that we can solve with our approach are ``nested''
recurrences, e.g., recurrences of the form $f(n)= g(f(f(n-1)))$.

We use the letters $x$, $y$, $z$ to denote variables, and $a$, $b$,
$c$, $d$ to denote constants and coefficients.  We use $f, g$ to
represent functions, and $e, t$ to represent arbitrary expressions. We
use $\varphi$ to represent arbitrary boolean constraints over a set of
variables. Sometimes, we also use $\beta$ to represent coefficients
obtained with linear regression. In all cases, the symbols can be
subscribed.
We use $\vec{x}$ to denote a finite sequence $\langle
x_1,x_2,\ldots,x_n\rangle$, for some $n>0$. Given a sequence $S$ and
an element $x$, $\langle x| S\rangle$ is a new sequence with first
element $x$ and tail $S$.

Given a piecewise function:
\begin{equation}
  \label{eq:1}
  f(\vec{x}) =  
  \begin{cases*}
    e_1(\vec{x}) & if $\varphi_1(\vec{x})$ \\
    e_2(\vec{x}) & if $\varphi_2(\vec{x})$ \\
    \vdots  & \vdots \\
    e_k(\vec{x}) & if $\varphi_k(\vec{x})$ \\
  \end{cases*}
\end{equation}
\noindent
where $f \in \mathcal{D}\to\mathbb{R}^{+}$, with
$\mathcal{D} = \{\vec{x} | \vec{x} \in \mathbb{Z}^{m} \wedge
\varphi_{\text{pre}}(\vec{x})\}$ for some boolean constraint $\varphi_{\text{pre}}$,
and $e_i(\vec{x}), \varphi_i(\vec{x})$ are arbitrary expressions and constraints over
$\vec{x}$ respectively. We say that
$\varphi_{\text{pre}}$ is the \emph{precondition} of $f$, and that $f$ is a
\emph{constrained recurrence relation} if and only if:

\begin{itemize}

\item $\exists i \in [1,k]$ such that $e_i$ contains a call to $f$.

\item $\exists i \in [1,k]$ such that $e_i$ does not contain any call
  to $f$ (i.e., it is in \emph{closed form}).

\item $\varphi_{\text{pre}} \models \bigvee\limits_{1\leq i \leq k}
  \varphi_i$.

\end{itemize}

Given a concrete input $\vec{d} \in \mathcal{D}$, we evaluate
$f(\vec{d})$ deterministically, assuming the evaluation of $f$ as a
nested \emph{if-then-else} control structure as follows:

\begin{center}
\begin{minipage}[h]{.3\linewidth}
  \begin{algorithmic}
    \IF {$\varphi_1(\vec{d})$} 
    \RETURN $e_1(\vec{d})$
    \ELSE
    \IF {$\varphi_2(\vec{d})$}
    \RETURN $e_2(\vec{d})$
    \ELSE
    \STATE $\cdots$
    \ENDIF
    \ENDIF 
  \end{algorithmic}
\end{minipage}
\end{center}
    
More formally,
let $\mathsf{def}(f)$ denote
the definition of a (piecewise) constrained recurrence relation $f$
represented as the sequence $\langle
(e_1(\vec{x}),\varphi_1(\vec{x})),\ldots,(e_k(\vec{x}),\varphi_k(\vec{x}))\rangle$,
where each element of the sequence is a pair representing a case.  The
order of such sequence determines the evaluation strategy.  Then, the
evaluation of
$f$ for a concrete value $\vec{d}$, denoted $\mathsf{EvalFun}(f(\vec{d}))$, is
defined as follows:
\begin{equation*}
  \mathsf{EvalFun}(f(\vec{d})) = \mathsf{EvalBody}(\mathsf{def}(f), \vec{d}) \\
\end{equation*}
\begin{equation*}
  \mathsf{EvalBody}(\langle (e,\varphi)|\mathsf{Ps}\rangle, \vec{d}) = 
  \begin{cases*}
    \eval{e}{\vec{d}}
    & if $\varphi(\vec{d})$ \\
    \mathsf{EvalBody}(\mathsf{Ps}, \vec{d}) & if $\neg \varphi(\vec{d})$ \\
  \end{cases*}
\end{equation*}

Our goal is to find a function $\hat{f} \in
\mathcal{D}\to\mathbb{R}^{+}$ such that
for all $\vec{d} \in \mathcal{D}$:
\begin{itemize}
\item $\text{ If }\mathsf{EvalFun}(f(\vec{d})) \text{ terminates, then }
  \mathsf{EvalFun}(f(\vec{d})) =
\eval{\hat{f}}{\vec{d}}$,
  and
\item $\hat{f}$ does not contain any recursive call in its definition.
\end{itemize}
    
In particular, we look for a definition of the form:
\begin{equation}
  \label{eq:2}
  \hat{f}(\vec{x}) = \beta_0 + \beta_1 \ t_1(\vec{x}) + \beta_2 \ t_2(\vec{x}) + \cdots +
  \beta_n \ t_n(\vec{x})   
\end{equation}
where $\beta_i \in \mathbb{R}$, and $t_i$ are expressions over $\vec{x}$,
not including recursive references to $\hat{f}$. If the above
conditions are met, we say that $\hat{f}$ is a \emph{closed form} for
$f$.

To illustrate the need of introducing an evaluation strategy for the
recurrence that is consistent with the termination of the program,
consider the following Prolog program which does not terminate for a
call \texttt{p(X)} where \texttt{X} is bound to an integer:

\begin{center}
\prettylstciao
  \begin{lstlisting}
p(X) :- X > 0, X1 is X + 1, p(X1).
p(X) :- X = 0.
  \end{lstlisting}
\end{center}

\noindent
The following
recurrence relation for its cost (in resolution steps) can be set up:
\begin{equation}
  \label{fig:cost-no-terminate}
  \begin{array}{rll}
    \stdub{\mathtt{p}}{x} = & 1 &  \text{ if } x = 0 \\ [-1mm]
    \stdub{\mathtt{p}}{x} = & 1 + \stdub{\mathtt{p}}{x+1} &  \text{ if } x > 0 \\ [-1mm]
  \end{array}
\end{equation}
\noindent
A CAS will give the closed form $\stdub{\mathtt{p}}{x} = 1 - x$ for
such recurrence, however, the cost analysis should give
$\stdub{\mathtt{p}}{x} = \infty$.

\paragraph{\textbf{Linear Regression}.}
Linear regression~\cite{Hastie-09a} is a statistical technique used to
approximate the linear relationship between a number of independent
variables and a dependent (output) variable. Given a vector of
independent (input) variables $X = (X_1,\ldots,X_p)^T \in \mathbb{R}^p$, 
we predict the output variable $Y$ using the formula
\begin{equation}
  \label{eq:2-2} Y = \beta_0 + \sum_{i=1}^p\beta_i X_i
\end{equation}
which is defined through the vector of coefficients $\beta =
(\beta_0,\ldots,\beta_p)^T \in \mathbb{R}^p$.
Such coefficients are estimated from a set of observations
$\{y_i,x_{i1},\ldots,x_{ip}\}_{i=1}^n$ so as to minimize a loss function,
most commonly the sum of squares
\begin{equation}
  \label{eq:4} \beta =
\underset{\beta \in \mathbb{R}^p}{\argmin}\sum_{i=1}^n\bigg(y_i - \beta_0 -
\sum_{j=1}^p x_{ij}\beta_j\bigg)^2
\end{equation}
Sometimes (as is our case) some of the input variables are not relevant to explain the output, but the above least-squares estimate will almost always assign nonzero values to all the coefficients. In order to force the estimate to make exactly zero the coefficients of irrelevant variables (hence removing them and doing \emph{feature selection}), various techniques have been proposed. The most widely used one is the Lasso~\cite{Hastie-15a}, which adds an $\ell_1$ penalty on $\beta$ (i.e., the sum of absolute values of each coefficient) to
Expression~\ref{eq:4}:
\begin{equation}
  \label{eq:4-1} \beta =
\underset{\beta \in \mathbb{R}^p}{\argmin}\sum_{i=1}^n\bigg(y_i - \beta_0 -
\sum_{j=1}^p x_{ij}\beta_j\bigg)^2 + \lambda\,\sum_{j=1}^p|\beta_j|
\end{equation}
where $\lambda \ge 0$ is a hyperparameter that determines the level
of penalization: the greater $\lambda$, the greater the number of
coefficients that are exactly equal to $0$. The Lasso has two advantages over other feature selection techniques for linear regression. First, it defines a convex problem whose unique solution can be efficiently computed even for datasets where either of $n$ or $p$ are large (almost as efficiently as a standard linear regression). Second, it has been shown in practice to be very good at estimating the relevant variables.

\section{Algorithmic Description of the Approach}
\label{sec:description-approach}

In this section we describe our approach for generating and checking
candidate solutions for recurrences that arise in resource analysis.
Algorithms~\ref{algo1} and~\ref{algo2} correspond to the
\emph{guesser} and \emph{checker} components, respectively, which are
shown in Figure~\ref{fig1}.

\begin{figure*}
  \centering
  \begin{algorithm}[H]
  \label{algo1}
    \SetKwInOut{Input}{Input}
    \SetKwInOut{Output}{Output}

    \Input{$F \in \mathcal{D} \to \mathbb{R}^+$: target recurrence relation.      
      \newline
      $\varphi_{\text{pre}}$: precondition defining $\mathcal{D}$.      
      \newline
      $T \subseteq \mathcal{D} \to \mathbb{R}^+$: set of base functions.
      \newline
      $\Lambda$: range of values to automatically choose a
      Lasso hyperparameter $\lambda \in \mathbb{R}^+$ that
      maximizes the performance of the model via cross-validation.
      \newline
      $k$: indicates performing $k-$fold cross-validation, $k \geq 2$. 
      \newline
      $\epsilon \in \mathbb{R}^+$: threshold for term ($t_i \in T$) selection.
    }
    \Output{$\hat{F} \in \mathsf{Exp}$: a candidate solution (or an approximation) for $F$. 
      \newline
      $S \in [0,1]$: score, indicates the accuracy of the estimation ($R^2$). 
    }
    
    $\mathcal{I} \gets \{\vec{x_i} | \vec{x_i} \in \mathbb{Z}^m \wedge \varphi_{\text{pre}}(\vec{x_i}) \}_{i=1}^N$ \tcp*{N Random inputs for F} \label{algo1:l1}
    $\mathcal{X} \gets  \{\langle F(\vec{x}) | E(T,\vec{x}) \rangle |
    \vec{x} \in \mathcal{I} \}$ \tcp*{Training set} \label{algo1:l2}
    $( \vec{\beta^{\prime}},\beta^{\prime}_0) \gets
    \mathsf{CVLassoRegression}(\mathcal{X}, \Lambda, k)$\;  \label{algo1:l3}
    $(T^{\prime},\mathcal{X}^{\prime}) \gets
    \mathsf{RemoveTerms}(T,\mathcal{X},\vec{\beta^{\prime}},\beta_0^{\prime},\epsilon)$\; \label{algo1:l4}
      $(\vec{\beta},\beta_0,S) \gets
      \mathsf{LinearRegression}(\mathcal{X}^{\prime})$\; \label{algo1:l5}
    $\hat{F} \gets \lambda \vec{x} \cdot \vec{\beta}^{T} \times
    E(T^{\prime},\vec{x}) + \beta_0$\; \label{algo1:l6}
    \Return{$(\hat{F}, S)$}\;
    \caption{Candidate Solution Generation (\emph{Guesser}).}
  \end{algorithm}
\end{figure*}
  
Algorithm~\ref{algo1} receives a recurrence relation for a function
$F$ to solve, a set of
base functions,
and a threshold to decide
when to discard irrelevant terms. The output is a closed-form
expression $\hat{F}$ for $F$, and a \emph{score} $S$ that reflects the
accuracy of the approximation, in the range $[0,1]$. If $S\sim1$, the
approximation can be considered a candidate solution. Otherwise,
$\hat{F}$ is a (possibly unsafe) approximation. In line~\ref{algo1:l1} we start by
generating a set $\mathcal{I}$ of random inputs for $F$. Each input $\vec{x_i}$ is a
$m$-tuple verifying precondition $\varphi_{\text{pre}}$, where $m$ is the
number of arguments of $F$. In line~\ref{algo1:l2} we produce the
training set $\mathcal{X}$.  The independent inputs are generated by evaluating the
base functions
in $T = \langle t_1, t_2, \ldots, t_p \rangle$ with each
tuple $\vec{x} \in \mathcal{I}$. This is done by using function $E$,
defined as follows:
\begin{equation*}
E(\langle t_1, t_2, \ldots, t_p \rangle, \vec{x}) = \langle t_1(\vec{x}), t_2(\vec{x}), \ldots, t_p(\vec{x}) \rangle
\end{equation*}
\noindent
We also evaluate the recurrence equation for input $\vec{x}$,
and add the observed output $F(\vec{x})$ as the first element in the
vectors of the training set. In line~\ref{algo1:l3} we generate a
first linear model by applying function $\mathsf{CVLassoRegression}$ to the
generated training set. $\mathsf{CVLassoRegression}$ performs a linear
regression with Lasso regularization. As already mentioned, Lasso
regularization requires a hyperparameter $\lambda$ that determines the
level of penalization for the coefficients. Instead of using a single
value for $\lambda$, $\mathsf{CVLassoRegression}$ uses a range of possible
values, applying cross-validation on top of the linear regression to
automatically select the best value for that parameter, from the given
range.
The parameter $k$ indicates performing $k-$fold cross-validation,
which means that the training set is split into $k$ parts or
\emph{folds}. Then, each fold is taken as the validation set, training
the model with the remaining $k-1$ folds. Finally, the performance
measure reported is the average of the values computed in the $k$
iterations.
The result of this function is the vector of coefficients
$\vec{\beta^{\prime}}$, together with
the intercept
$\beta_0^{\prime}$. These coefficients are used in line~\ref{algo1:l4}
to decide which base functions
are discarded before the last regression
step. Note that $\mathsf{RemoveTerms}$
removes the base functions
from $T$
together with their corresponding input values from the training set
$\mathcal{X}$, returning the new set of base functions
$T^{\prime}$
and its corresponding training set $\mathcal{X}^{\prime}$. In
line~\ref{algo1:l5}, standard linear regression (without
regularization nor cross-validation) is applied, obtaining the final
coefficients $\vec{\beta}$ and $\beta_0$. Additionally, from this step
we also obtain the score $S$ of the resulting model. In
line~\ref{algo1:l6} we set up the resulting closed-form expression,
given as a function on the variables in $\vec{x}$. Note that we use
the function $E$ to bind the variables in the base functions
to the arguments of the closed-form expression. Finally, the
closed-form expression and its corresponding score are returned as the
result of the algorithm.

Algorithm~\ref{algo2}
mainly relies on an SMT-solver and a CAS.
Concretely, given the
constrained recurrence relation $F \in \mathcal{D} \to \mathbb{R}^+$ defined as
\begin{equation*}
  \label{eq:1-1}
  F(\vec{x}) =
  \begin{cases*}
    e_1(\vec{x}) & if $\varphi_1(\vec{x})$ \\
    e_2(\vec{x}) & if $\varphi_2(\vec{x})$ \\
    \vdots  & \vdots \\
    e_k(\vec{x}) & if $\varphi_k(\vec{x})$ \\
  \end{cases*}
\end{equation*}
our algorithm constructs the logic formula: 
\begin{equation}
  \label{smtrep}
  \bigg \llbracket \bigwedge\limits_{i=1}^k \left(\left(\bigwedge\limits_{j=1}^{i-1} \neg  \varphi_j(\vec{x}) \right)  \wedge \varphi_i(\vec{x}) \wedge \varphi_{\text{pre}}(\vec{x}) \implies \mathsf{Eq}_i \right) \bigg \rrbracket_{\text{SMT}}
\end{equation}
where $\mathsf{Eq}_i$ is the result of replacing in $F(\vec{x}) = e_i(\vec{x})$
each occurrence of $F$,
if possible, by the definition of the candidate solution $\hat{F}$ (by
using $\mathsf{replaceCalls}$ in line~\ref{linereplacecalls}), and performing a
simplification by the CAS (by using $\mathsf{simplifyCAS}$ in
line~\ref{linesimpl}). A goal of such simplification is to obtain
(sub)expressions supported by the SMT-solver.
The function
$\mathsf{replaceCalls}(\mathsf{expr},F(\vec{x}^\prime),\hat{F},\varphi_{\text{pre}},\varphi)$
replaces
every subexpression in $\mathsf{expr}$ of the form $F(\vec{x}^\prime)$ by
$\hat{F}(\vec{x}^\prime)$, if $\varphi_{\text{pre}}(\vec{x}^\prime) \wedge \varphi
\implies \varphi_{\text{pre}}(\vec{x}^\prime)$.
The operation
$\llbracket  e \rrbracket_{\text{SMT}}$
is the translation of any expression $e$ to an
SMT-LIB
expression. Although all variables appearing in Formula~\ref{smtrep} are
declared as integers, we omit these details in Algorithm~\ref{algo2}
and in Formula~\ref{smtrep} for the sake of brevity.
Note that this encoding is consistent with the evaluation ($\mathsf{EvalFun}$)
described in Section~\ref{sec:prelim}. Finally, the algorithm asks the
SMT-solver
for models of the negated formula (line~\ref{lastline}). If no model
exists, then it returns $\mathsf{true}$, concluding that $\hat{F}$ is an exact
solution to the recurrence, i.e., $\hat{F}(\vec{x}) = F(\vec{x})$ for
any input $\vec{x} \in \mathcal{D}$ such that $\mathsf{EvalFun}(F(\vec{x}))$
terminates. Otherwise, it returns $\mathsf{false}$. Note that, if it is not
possible to replace all occurrences of $F$ by $\hat{F}$, or if after
performing the simplification by $\mathsf{simplifyCAS}$ there are subexpressions
not supported by the SMT-solver,
then the algorithm finishes returning $\mathsf{false}$.

\begin{figure*}[t]
  \centering
  \begin{algorithm}[H]
  \label{algo2}
    \SetKwInOut{Input}{Input}
    \SetKwInOut{Output}{Output}

    \Input{$F \in \mathcal{D} \to \mathbb{R}^+$: target recurrence relation.
      \newline
      $\varphi_{\text{pre}}$: precondition defining $\mathcal{D}$.
      \newline
      $\hat{F} \in \mathsf{Exp}$: a candidate solution for $F$.
    }
    \Output{$\mathsf{true}$ if $\hat{F}$ is a solution for $F$, $\mathsf{false}$ otherwise. 
    }
    $\varphi_{\text{previous}} \gets \mathsf{true}$ \; $\mathsf{Formula} \gets \mathsf{true}$ \;
    \ForEach{$(e,\varphi) \in \mathsf{def}(F)$}{
      $\mathsf{Eq} \gets \mathsf{replaceCalls}(``F(\vec{x}) - e = 0",F(\vec{x}),\hat{F},\varphi_{\text{pre}},\varphi)$\; \label{linereplacecalls}
      \eIf{$\neg \: \mathsf{containsCalls}(\mathsf{Eq},F)$}{
      $\mathsf{Eq} \gets \mathsf{simplifyCAS}(\mathsf{inlineCalls}(\mathsf{Eq},\hat{F},\mathsf{def}(\hat{F})))$\; \label{linesimpl}
      \eIf{$\mathsf{supportedSMT}(\mathsf{Eq})$}{
        $\mathsf{Formula} \gets ``\mathsf{Formula} \wedge (\varphi_{\text{pre}} \wedge \varphi_{\text{previous}} \wedge \varphi \implies \mathsf{Eq})"$\;
        $\varphi_{\text{previous}} \gets ``\varphi_{\text{previous}} \wedge \neg \varphi"$ \;
      }{
        \Return{$\mathsf{false}$}\;
      }}{
      \Return{$\mathsf{false}$}\;
    }
    
      }
    
    \Return{$(\not\models_{\text{SMT}} \llbracket \neg \mathsf{Formula} \rrbracket_{\text{SMT}})$}\; \label{lastline}
    \caption{Solution Checking (\emph{Checker}).}
  \end{algorithm}
\end{figure*}


\section{Our Approach in the Context of Static Cost Analysis}
\label{sec:application-cost-analysis}

In this section, we describe how our approach could be used in the
context of the motivating application, Static Cost Analysis.  Although
it is general, and could be integrated into any
cost analysis system based on recurrence solving, we illustrate
its use in the context of the \ciaopp system.
Using a logic program, we first illustrate how
\ciaopp
sets up recurrence relations representing the sizes of output
arguments of predicates and the cost of such predicates.  Then, we
show how our novel approach is used to solve a recurrence relation
that cannot be solved by \ciaopp.

\begin{example}
\label{ex:running-code-regr}
Consider predicate \verb-p/2-
in Figure~\ref{fig:example1-code}, and
calls to it where the first argument is bound to a non-negative
integer
and the second one is a free variable. Upon success of these calls,
the second argument is bound to an non-negative integer too.
Such calling mode, where the first argument is input and the second
one is output, is automatically inferred by \ciaopp
(see~\cite{ciaopp-sas03-journal-scp-short} and its references).
\end{example}

\begin{figure}[h]
  \centering
  \prettylstciao
  \begin{lstlisting}
:- entry p/2: nnegint*var.
p(X,0):-
  X=0.
p(X,Y):-
  X>0,
  X1 is X - 1,
  p(X1,Y1),
  p(Y1,Y2),
  Y is Y2 + 1.
\end{lstlisting}
  \caption{A program with a nested recursion.}
  \label{fig:example1-code}
\end{figure}

The \ciaopp{} system first infers size relations for the different
arguments of predicates, using a rich set of size metrics
(see~\cite{resource-iclp07-short,plai-resources-iclp14-short} for details). Assume
that the size metric used in this example, for the numeric argument
\texttt{X} is the \emph{actual value} of it (denoted \texttt{int(X)}).
The system will try to infer a function ${\tt S}_\mathtt{p}(x)$ that
gives the size of the output argument of \texttt{p/2} (the second
one), as a function of the size ($x$) of the input argument (the first
one). For this purpose, the following size relations for ${\tt
  S}_\mathtt{p}(x)$ are automatically set up (the same as the
recurrence in Equation~\ref{eq:recurrence} used in
Section~\ref{sec:overview} as example):
\begin{equation}
  \label{fig:size1}
  \begin{array}{rll}
    \sizeub{\mathtt{p}}{x} = & 0 &  \text{ if } x = 0 \\ [-1mm]
    \sizeub{\mathtt{p}}{x} = & \sizeub{\mathtt{p}}{\sizeub{\mathtt{p}}{x-1}}  + 1 &  \text{ if } x > 0 \\ [-1mm]
  \end{array}
\end{equation}
\noindent
The first and second recurrence correspond to the first and second
clauses respectively (i.e., base and recursive cases). Once recurrence
relations (either representing the size of terms, as the ones above,
or the computational cost of predicates, as the ones that we will see
latter) have been set up, a solving process is started.

Nested recurrences, as the one that arise in this example, cannot be
handled by most state-of-the-art recurrence solvers. In particular,
the modular solver used by \ciaopp{} fails to find a closed-form
function for the recurrence relation above. In contrast, the novel
approach that we propose, sketched in next section, obtains the
closed form $\hat{{\tt S}}_\mathtt{p}(x) = x$, which is an
exact solution of such recurrence (as shown in Section~\ref{sec:overview}).

Once the size relations have been inferred, \ciaopp{} uses them
to infer the computational cost of a call to \texttt{p/2}. For
simplicity, assume that in this example, such cost is given in terms
of the number of \emph{resolution steps},
as a function of the size of the input argument, but note that
\ciaopp's cost analysis is parametric with respect to resources, which
can be defined by the user by means of a rich assertion language, so
that it can infer a wide range of resources, besides resolution steps.
Also for simplicity, we assume that
all builtin predicates, such as arithmetic/comparison operators
have zero cost (in practice there is a ``trust''assertion for each
builtin that specifies its cost as if it had been inferred by the
analysis).

In order to infer the cost of a call to \texttt{p/2}, represented as
$\stdub{\mathtt{p}}{x}$, \ciaopp{} sets up the following
\costrelations{}, by using the size relations inferred previously:
\begin{equation}
  \label{fig:cost1}
  \begin{array}{rll}
    \stdub{\mathtt{p}}{x} = & 1 &  \text{ if } x = 0 \\ [-1mm]
    \stdub{\mathtt{p}}{x} = & \stdub{\mathtt{p}}{x-1} + \stdub{\mathtt{p}}{\sizeub{\mathtt{p}}{x-1}}  + 1 &  \text{ if } x > 0 \\ [-1mm]
  \end{array}
\end{equation}
\noindent
We can see that the cost
of the second recursive call to predicate \texttt{p/2} depends on the
size of the output argument of the first recursive call to such
predicate, which is given by function $\sizeub{\mathtt{p}}{x}$, whose
closed form $\sizeub{\mathtt{p}}{x} = x$ is computed by our approach,
as already explained.
Plugin such closed form into the
recurrence relation above, it can be solved now by \ciaopp, obtaining
$\stdub{\mathtt{p}}{x} = 2^{x+1} - 1$.


\section{Implementation and Experimental Evaluation}
\label{sec:implem-and-eval}

We have implemented a prototype of our novel approach
and performed an experimental evaluation in the context of the \ciaopp
system, by solving
recurrences generated during static cost analysis.
Our prototype takes a recurrence and returns a closed form obtained
together with two measures: 1) the accuracy of the estimation
(\emph{score}) of the candidate closed-form solution generated by the
machine learning phase, and 2) an indication of whether such closed form is
an exact solution of the recurrence (i.e., if it has been formally
verified).
It is implemented in \emph{Python 3}, using
\emph{Sympy}~\cite{sympy} as CAS, and
\emph{Scikit-Learn}~\cite{scikit-learn} for the regression with
Lasso regularization. We use \emph{Z3}~\cite{z3} as
SMT-solver, and \emph{Z3Py}~\cite{z3py} as interface.


\begin{table}[t]
\centering
\caption{Closed forms obtained with the previous
  (\textbf{CF}) and new solver (\textbf{CFNew}).}\label{mlcost:tab-exp1}
\begin{tabular}{l|l|p{0.8cm}|l|c}  
\hline\hline  
\textbf{Bench} & \textbf{Recurrence} & \textbf{CF} & \textbf{CFNew} &
\textbf{T (s)}
\\ \midrule
merge-sz
& \scalebox{0.9}{$f(x,y) =
    \begin{cases*}
      max(f(x-1,y), \\ f(x,y-1)) + 1 & if $x > 0 \land y > 0$ \\
      x                              & if $x > 0 \land y = 0$ \\
      y                              & if $x = 0 \land y > 0$ \\
    \end{cases*}$}
& $-$
& $x + y$
& 0.92 
\\ \hline
merge
& \scalebox{0.9}{$f(x,y) =
    \begin{cases*}
      max(f(x-1,y), \\ f(x,y-1)) + 1 & if $x > 0 \land y > 0$ \\
      0                              & if $x = 0 \lor  y = 0$  \\
    \end{cases*}$}
& $-$
& \scalebox{0.9}{$
    \begin{cases*}  
      x+y-1 & if $x > 0 \land y > 0$ \\
      0     & if $x = 0 \lor  y = 0$  \\
    \end{cases*}$}       
& 0.71
\\ \hline
nested
& \scalebox{0.9}{$f(x) =
  \begin{cases*}
    f(f(x-1)) + 1 & if $x > 0$ \\
    0             & if $x = 0$ \\
  \end{cases*}$}
& $-$
& $x$
& 0.13 
\\ \hline
open-zip
& \scalebox{0.9}{$f(x,y) =
    \begin{cases*}
      f(x-1,y-1) + 1  & if $x > 0 \land y > 0$ \\
      f(x,y-1)   + 1  & if $x = 0 \land y > 0$ \\
      f(x-1,y)   + 1  & if $x > 0 \land y = 0$ \\
      0               & if $x = 0 \land y = 0$ \\
    \end{cases*}$}
& $-$
& $\max\left(x, y\right)$
& 0.12 
\\ \hline
div
& \scalebox{0.9}{$f(x,y) =
    \begin{cases*}
      f(x-y,y) + 1 & if $x >= y \land y > 0$ \\
      0            & if $x <  y \land y > 0$ \\
    \end{cases*}$}
& $-$
& $\left\lfloor{\frac{x}{y}}\right\rfloor$
& 0.13 
\\ \hline
div-ceil
& \scalebox{0.9}{$f(x,y) =
    \begin{cases*}
      f(x-y,y) + 1 & if $x >= y \land y > 0$ \\
      1            & if $x < y  \land x > 0$ \\
      0            & if $x = 0  \land y > 0$ \\
    \end{cases*}$}
& $-$
& $\left\lceil{\frac{x}{y}}\right\rceil$
& 0.12 
\\ \hline
s-max
& \scalebox{0.9}{$f(x,y) =
    \begin{cases*}
      max(y,f(x-1,y)) + 1 & if $x > 0$ \\
      y                   & if $x = 0$ \\
    \end{cases*}$}
& $x + y$
& $x + y$
& 0.12 
\\ \hline
s-max-1
& \scalebox{0.9}{$f(x,y) =
    \begin{cases*}
      max(y,f(x-1,y+1)) + 1 & if $x > 0$ \\
      y                     & if $x = 0$ \\
    \end{cases*}$}
& $-$
& $2x + y$
& 0.14
\\ \hline
sum-osc
& \scalebox{0.9}{$f(x,y) =
    \begin{cases*}
      f(x-1,y)   + 1 & if $x > 0 \land y > 0$ \\
      f(x+1,y-1) + y & if $x = 0 \land y > 0$ \\
      1              & if $y = 0$ \\
    \end{cases*}$}
& $-$
& \scalebox{0.9}{$
    \begin{cases*}
      x + \frac{y^{2}}{2} + \frac{3 y}{2} & if $y > 0$ \\
      1                                   & if $y = 0$ \\
    \end{cases*}$}
& 0.13 
\\ \hline\hline  
\end{tabular}
\end{table}


Our experimental results are shown in Table~\ref{mlcost:tab-exp1}.
Column \textbf{Bench} shows the name that we have assigned to each
recurrence that we have chosen (which is inspired in the logic program such
recurrence originated from during cost/size analysis), and Column
\textbf{Recurrence} shows their definitions, where we use the same
function symbol, $f$, for all of them.  Such recurrences are
challenging for \ciaopp,
either because they cannot be
solved by any of the back-end solvers, or because they are necessarily
over-estimated in the solving process. Some recurrences, like
\textbf{nested}, are problematic even for most of the current
state-of-the-art solvers.
For each recurrence and for each argument $x$ of it, there is an
implicit constraint that $x$ is an integer, $x \geq 0$, which we do
not include for brevity. Also, the disjunction of all the constraints
defining the cases of the recurrence is a constraint $\varphi_{\text{pre}}$
that defines the domain of the corresponding function.
Column \textbf{CF} shows the closed forms
obtained by our previous recurrence solver, and Column \textbf{CFNew}
shows the closed forms
obtained by our approach,
applying Algorithms~\ref{algo1} and~\ref{algo2}.
All of them have been verified as exact solutions to the recurrences
by Algorithm~\ref{algo2}.
As already said in Section~\ref{sec:prelim}, the closed form solution
of any recurrence gives the same results as the recurrence for the
inputs for which the evaluation of the recurrence terminates.  The
base cases of the recurrences have been considered apart from the
others, and included in the final solution.
Finally, Column \textbf{T(s)} shows the total time, in seconds
(executing on a MacBook Pro machine, 2.4GHz Intel Core i7 CPU, 8 GB
1333 MHz DDR3 memory), needed to obtain the closed forms and
verify
them.
For all the experiments, we have set $k = 2$, in order to perform
$2-$fold cross-validation. We have also set the range for $\lambda$ to
$100$
values taken from the interval
$[0.001, 1]$, and $\epsilon = 0.05$. Regarding the set $T$ of base functions,
for recurrences
with one or two arguments, we provide a predefined set of
representative functions of the most common complexity orders, as well
as some compositions of them. For recurrences with three or more
arguments, we provide an initial set of simple functions, that are
combined automatically to generate the base functions $t_i$ for the
set $T$.

As we can see, none of the recurrences are solvable by the current
\ciaopp solver, except \texttt{s-max}.  The specialized solver for
such recurrence has been developed relatively recently. Also, none of
the recurrences are solvable by the CASs
\emph{Mathematica}~\cite{mathematica-v13-2} and
\emph{Sympy}~\cite{sympy}, which we can arguably consider
state-of-the-art CASs.
In contrast, our new solver is able to infer exact closed-forms
functions for all the recurrences in a reasonable time.

\section{Conclusions and Future Work}
\label{sec:conclusions}
We have developed a novel
approach for solving or approximating arbitrary, constrained
recurrence relations.  It consists of a \emph{guess} stage that uses a
sparse linear regression via Lasso regularization and cross-validation
to infer a candidate closed-form solution, and a \emph{check} stage
that combines an SMT-solver and a CAS to verify that such candidate is
actually a solution.  We have implemented a prototype and evaluated it
with recurrences generated by the cost analysis module of \ciaopp, and
are not solvable by it nor by the (arguably state-of-the-art) CASs
\emph{Mathematica} and \emph{Sympy}.  The experimental results are
quite promising, showing that our approach can find exact, verified,
closed-form solutions, in a reasonable time, for such recurrences,
which imply potentially, arbitrarily large accuracy gains in cost
analysis of (logic) programs.
Not being able to solve a recurrence can cause huge accuracy losses,
for instance, if such a recurrence corresponds to a predicate that is
deep
in the control flow graph of the program, and such accuracy loss is
propagated to the
main predicate, inferring not useful information at all.

Since our technique uses linear regression with a randomly generated
training set (by evaluating the recurrence to obtain the dependent
value), it is not guaranteed that a solution can be found. Even if an
exact
solution is found in the first stage, it is not always possible to
prove its correctness in the second stage. Therefore, in this sense,
this approach is \emph{not complete}. However, it is able to find some
solutions that current state-of-the-art solvers are unable to find.
As a proof of concept, we have considered a particular deterministic
evaluation for constrained recurrence relations, and the verification
of the candidate solution is consistent with this evaluation. However,
it is possible to implement different evaluation semantics for the
recurrences, adapting the verification stage accordingly. Note that we
need to
require the termination of
the recurrence evaluation as a precondition
for the conclusions obtained. This is also due to the particular
evaluation strategy of recurrences that we are considering. In
practice, non-terminating recurrences can be discarded in the first
stage, by setting a timeout.
Our approach can also be combined with a termination prover in order
to guarantee such a precondition.  Finally, note that an alternative use
of our tool is to omit the verification stage, using only the
closed-form function inferred by the first stage, together with an
error measure. This can be useful in some applications (e.g.,
granularity control in parallel/distributed computing) where it is
enough to have good although unsafe approximations.

As a future work, we plan to fully integrate our novel solver into the
\ciaopp system, combining it with its current set of back-end solvers in order
to improve the static cost analysis. We also plan to further refine and
improve our
algorithms in several directions. As already
explained, currently the set $T$ of base functions is fixed,
user-provided.  We plan to automatically infer it by using different
heuristics.
We can perform an automatic analysis of
the recurrence we are solving, to extract some features that allow
selection of the terms that most likely are part of the solution. For
example, if the recurrence has a nested, double recursion, then we can
select a quadratic term, etc. Also, machine learning techniques may be
applied to learn a good set of base functions from some features of
the programs.

\begin{small}
\paragraph{\textbf{Acknowledgments}}

This work has been partially supported by MICINN projects
PID2019-108528RB-C21 \emph{ProCode}, TED2021-132464B-I00
\emph{PRODIGY}, and FJC2021-047102-I, and the Tezos foundation.  The
authors would also like to thank Louis Rustenholz, John Gallagher,
Manuel Hermenegildo, Jos\'{e} F. Morales and the anonymous reviewers
for very useful feedback.  Louis Rustenholz also recreated the
experimental results and double-checked them.
    
\end{small}

\bibliographystyle{eptcs}

\end{document}